\documentclass[12pt]{article}
\usepackage{graphicx}
\usepackage{amssymb,amsmath,amsfonts,palatino,amsthm}
\usepackage{amssymb}
\usepackage{epstopdf}
\usepackage{slashed}
\newcommand{\f}{\begin{equation}}
\newcommand{\ff}{\end{equation}}

\begin{document}

\title{Cosmological signatures of time-asymmetric gravity}
\author{Marina Cort\^{e}s${}^{1,2,3}$, Andrew R.~Liddle${}^{1,2}$ and 
Lee Smolin${}^{2,4}$ 
\\
\\
${}^{1}$Institute for Astronomy, University of Edinburgh \\
Blackford Hill, Edinburgh EH9 3HJ, United Kingdom
\\
\\
${}^{2}$Perimeter Institute for Theoretical Physics\\
31 Caroline Street North, Waterloo, Ontario N2J 2Y5, Canada
\\
\\
${}^{3}$Centro de Astronomia e Astrof\'{i}sica da Universidade de Lisboa\\
Faculdade de Ci\^encias, Edif\'{i}cio C8, Campo Grande, 1769-016 Lisboa, Portugal
\\
\\
${}^{4}$Department of Physics and Astronomy, University of Waterloo
}
\date{\today}

\maketitle

\begin{abstract}
We develop the model proposed by Cort\^es, Gomes \& Smolin \cite{CGS} to predict cosmological signatures of time-asymmetric extensions of general relativity. 
Within this class of models the equation of motion of chiral fermions is modified by a torsion term. This term leads to a dispersion law for neutrinos that associates a new time-varying energy with each particle. We find a new neutrino contribution to the Friedmann equation resulting from the torsion term in the Ashtekar connection. In this note we explore the phenomenology of this term and observational consequences for cosmological evolution. We show that constraints on the critical energy density will ordinarily render this term unobservably small, a maximum of order $10^{-25}$ of the neutrino energy density today. However, if the time-asymmetric dark energy is tuned to cancel the cosmological constant, the torsion effect may be a dark matter candidate.
\end{abstract}

\newpage

\tableofcontents

\section{Introduction}

Two of us, along with Gomes, recently proposed a simple extension of general relativity with explicit breaking of time-reversal symmetry \cite{CGS}. Such theories give an explicit directionality to time, while retaining local energy-momentum conservation as time-translation invariance remains unbroken. The extension is expressed in terms of the Hamiltonian formulation of general relativity.  The Hamiltonian constraint is modified by the addition of a  new term which is linear in $\pi$, the trace of the field conjugate to the metric in a Hamiltonian formalism.  In the extension, $\pi$ is multiplied by a new free function $f(V)$ of the volume $V$ of slices in the constant mean curvature (CMC) gauge. The extension is designed to preserve the first-class nature of the constraints.  Since $f(V)$ is a free function we really have a class of extensions to general relativity.  

The restriction to CMC gauge signals that the consistency of this proposal rests on the existence of shape dynamics, which is a consistent reformulation of general relativity in which refoliation invariance is replaced by a local three-dimensional
scale invariance \cite{SD1}-\cite{Flavio_tutorial}.  This is consistent with the reduction to FRW cosmologies, where the homogeneous spatial slices coincide with a CMC foliation.

In cosmology, the new term appears in the Friedmann equation, and the freedom in selecting $f(V)$ permits a wide range of possible behaviours; Ref.~\cite{CGS} considered three cases, which they called time-asymmetric dark energy, dark curvature, and dark radiation. Current observations are able to limit, but not exclude, such contributions to the Friedmann equation.  A more extensive analysis of background dynamics in this scenario was given in Ref.~\cite{LS}.

Reference~\cite{CGS} also showed that the new term would modify the propagation equations for chiral fermions via a new torsion term in the Weyl equation. The purpose of this note is to further explore the possible observational consequences of this term for cosmological evolution. In the following we focus on neutrinos but our considerations apply to any fermions that may be considered purely chiral during the cosmological era in question.

The results obtained here show that the program of research described in Refs.~\cite{TR}-\cite{ECS2}, based on the hypotheses that time is fundamental and irreversible, can have concrete implications for observational cosmology. Other applications of torsion to cosmology are described in Ref.~\cite{other}.

\section{Cosmological evolution in time-asymmetric general relativity}

Reference~\cite{CGS} showed that the introduction of the time-asymmetric term to the Hamiltonian results in an addition to the Friedmann equation of a term proportional to $g^2(a)$ where $g(a)$ is related to the free function $f(V)$ mentioned above above by $g = af/G$, $G$ is Newton's constant and $a$ is the scale factor. That is, the Friedmann equation becomes
\begin{equation}
\left(\frac{\dot{a}}{a}\right)^2 = \frac{\Lambda}{3} - \frac{K}{a^2} + \frac{8\pi G}{3} \, \rho + \frac{G^2 g^2}{a^2} 
\end{equation}
where $\Lambda$, $K$ and $\rho$ are the usual (bare) cosmological constant, curvature, and energy density, and the last term is new. Our ability to constrain this is rather limited, since the arbitrariness of the function $g(a)$ means that any cosmological history $a(t)$ could be obtained, but nevertheless we can propose possible behaviours through making particular ansatzes for $g$. The most interesting of these is time-asymmetric dark energy, where $g \propto a$~\cite{CGS}.

This is not however the whole story, because the same modification to the gravitational Hamiltonian affects the propagation of chiral fermions in a non-trivial way. As derived in Ref.~\cite{CGS}, the new term gives rise to an Ashtekar connection which differs from the usual self-dual connection of space--time by a torsion term. This implies that there are two geometries: apart from the metric governing space--time geometry and the propagation of photons, there is another chiral connection which governs the propagation of chiral fermions. The difference between this chiral connection and the left-handed part of the metric connection can be described as a torsion.  The modified fermion propagation alters the form of the fermionic energy density, which is to appear in the Friedmann equation. Since this `torsion' effect is determined by the same function $g(a)$, its form is predicted once the modification to the Friedmann equation is specified, raising the possibility of a directly-observable effect.

We work within the Hamiltonian framework, as that is the language in which the time-asymmetric theories we work with are expressed.
We can derive the Weyl equation of motion for a chiral spin-$\frac{1}{2}$ fermion $\Psi_B$ by noting that a chiral fermion is added to the 
system by adding a term to each of the Hamiltonian, diffeomorphism, and gauge constraints \cite{CGS} .  These are\footnote{Here 
$\mu, \nu , \dots = 0,1,2,3$ are manifold space--time indices, while $\alpha, \beta , \dots = 1,2,3 $   are manifold indices on the spatial slice.  $a,b = 0,1,2,3$ are internal space--time indices that label the tetrads, while $i,j = 1,2,3$ are spatial internal indices or triad labels.  $A,B = 0,1$ are left-handed two-component spinor indices.}
\f
{\cal C}^{\Psi} =  \Pi^A e^a_i \sigma^{ i \ \ B}_A D_a \Psi_B
\label{hampsi}
\ff
\f
{\cal D}_a^{\Psi} =  \Pi^A D_a \Psi_A
\ff
\f
{\cal J}^{AB}= \Pi^{(A} B^{B)}
\ff
Here $\Pi^B$ is the momentum conjugate to $\Psi_B$ and $\sigma^{ i \ \ B}_A$ are the usual Pauli matrices.
$D_a \Psi_A$ is the left-handed space--time connection which, very importantly, 
{\it depends on $\pi^{ab}$} (the momentum conjugate to the metric $g_{ab}$ on the spatial slices)
\f
D_a \Psi_A = \partial_a \Psi_A + A_{a A}^{ \  B} \Psi_B
\ff 
where $A_{a A}^{\  B} = A_{a }^{ \  i} \sigma_A^{i \ B}$ is the  Ashtekar
connection.  It is defined by \cite{AA}
\f
A^{\ i}_a = \Gamma_a^i +\frac{\imath G}{\sqrt{g}} ( \pi_a^i -e_a^i \pi  )
\label{Aeq}
\ff
where  $\Gamma_{a }^{\  i} $ is the Christoffel three-connection and the tildes 
indicate densities.\footnote{We remain in the extended phase space \cite{AA}, whose canonical coordinates are
$e_a^i$ and $\tilde{\pi}^b_j$, which have the canonical commutation relations,   
$ \{ e_a^i (x), \tilde{\pi}^b_j (y) \} = \delta^3 (x,y) \delta^b_a \delta^i_j $.}

The Weyl equation of motion for a chiral spin-$\frac{1}{2}$ fermion is obtained from Eq.~(\ref{hampsi})
\f
\dot{\Psi}_A =  \{ \Psi_A , {\cal H}^{\Psi} (N) \}= Ne^a_i \sigma^{ i \ \ B}_A D_a \Psi_B 
\ff
The time derivative is uncorrected because we are in $A_0=0$ gauge.
We then have the Weyl equation,
\f
\slashed{\cal D}_{A}^B \Psi_B  =\frac{\partial \Psi_A}{\partial t} - e^a_i \sigma^{ i \ \ B}_A D_a \Psi_B =0
\label{Weyl}
\ff

To work out the implications of this for cosmology, it is enough to start with the phase-space action which defines the reduction to FRW cosmologies.
\f
S=  v_0\int dt \left (
\pi \dot{a} - N_{\rm lapse} {\cal C}
\right )
\label{FRWaction}
\ff
To get this we have reduced the canonical variables via
\f
g_{ab}= a^2 (t) q^0_{ab}
\label{FRW1}
\ff
where $q^0_{ab}$ is a non-dynamical metric which is flat or constantly curved, and
\f
\tilde{\pi}^{ab} =  \frac{1}{3a} \sqrt{q^0} q^{ab}_0  \pi (t)
\label{FRW2}
\ff
The fiducial volume of the universe is
\f
v_0 = \int_\Sigma \sqrt{q^0}
\ff

The Hamiltonian constraint, with the homogeneous lapse $N_{\rm lapse}=1$,   generates time reparametrizations,
\f
{\cal C} = \frac{G}{2a}   \pi^2 + G g(a) \pi - a^3 V(a)
\ff
where $g(a)= af(a)/G $ is a function of $a$. The potential $V$ is
\f
V=  \frac{\Lambda}{6G} - \frac{k}{2G a^2} + \frac{4 \pi  \rho_0 }{3a^3 } 
\ff
We assume that the effect of the fermion term,  Eq.~(\ref{hampsi}), has been absorbed into the potential, as we describe below.

We vary first by $\pi$ to find
\f
\frac{\dot{a}}{N_{\rm lapse}} = G \left ( \frac{\pi}{a}  + g \right )
\label{aeom}
\ff
This gives us
\f
\pi =  \frac{a^2}{N_{\rm lapse} G }H - a g
\label{pieq}
\ff
in terms of the usual Hubble parameter, $H=\dot{a}/a$.
If we vary the action by the Lagrange multiplier, $N_{\rm lapse}$, 
we find the 
modified Friedmann equation from setting the Hamiltonian ${\cal H}=0$, yielding
\f
{\cal H}=  a^3 \left ( \frac{1}{2 N_{\rm lapse}^2 G } H^2 - \frac{G g^2}{2a^2} - V 
\right ) =0 
\label{C}
\ff
while varying by $a$ gives an equation for $\dot{\pi}$ 
\f
\frac{1}{N} \dot{\pi} =  \frac{G \pi^2}{2a^2} + 3 a^2 V- a^3 V^\prime -G g^\prime \pi
\label{pieom}
\ff
where primes are derivatives with respect to $a$.

To find the Ashtekar connection for FRW spacetimes, we choose a gauge where
\f
e^0_\mu = \delta|^0_\mu  , \ \ \ e^i_\mu = \frac{1}{a} \delta^i_\mu
\ff
Using Eqs.~(\ref{Aeq}) and (\ref{pieq}) the Ashtekar connection is 
\f
A_\alpha ^i = \Gamma_\alpha ^i - \frac{2 \imath G}{3} \delta_\alpha^i \left(\frac{aH}{N_{\rm lapse} G} - g(a) \right)
\ff
For flat FRW the Christoffel term would vanish, with the first term in brackets being the usual GR Ashtekar connection and the final term being the new contribution from the time-asymmetry.
This results in an equation of motion
\f
\slashed{\cal D}_{A}^B \Psi_B  =0 \rightarrow 
 \slashed{\cal D}_{A}^{0 B} \Psi_B   -  \imath \mu  \Psi_A =0 
\ff
where  $ \slashed{\cal D}_{A}^{0 B} \Psi_B =0$ is the standard Weyl equation (\ref{Weyl}) and $\mu$ is a frequency coming from the $g(a)$ dependence in the canonical momentum, which works out to be
\f
\mu (a) =  2 G\frac{g(a)}{a}
\ff
This results in a shift in the frequency--wavevector relation 
\f
\omega = k +  \mu (a) 
\ff
where $k$ is the modulus of the wavevector ${\bf k}$. Note that this is different from the effect of a rest-mass term, which would lead to $\omega^2 = k^2 + m^2$.

Consider a set of particle species in thermal equilibrium with present temperature $T_0$, which for neutrinos is approximately 2 Kelvin. The energy per particle in thermal equilibrium is $E \equiv \hbar \omega \simeq 3k_{\rm B} T$, and the Fermi--Dirac distribution indicates a number density for each species of $n \simeq 0.1 T^3$ (where here and henceforth we adopt naturalized units $\hbar = c = k_{\rm B} = 1$, while retaining $G \equiv m_{\rm Pl}^{-2}$ in order to facilitate dimensional analysis). If there are $N_{\rm f}$ degrees of freedom (e.g.\ 2 spin states for each of three neutrino flavours), and the present average wavenumber is $k_0$ so that the physical wavenumber at a general epoch is $k_0/a$, the result is that the total energy density, as a function of scale factor $a$, is
\f
\rho = \frac{n_0 N_{\rm f}}{a^3} \times \mbox{mean energy per particle} =\frac{n_0 N_{\rm f}}{a^3} \left( \frac{k_0}{a} + 2 G\frac{g(a)}{a} \right) 
\ff

We have to add the new torsion-dependent term to the potential energy. Hence the whole potential is 
\f
V=  \frac{\Lambda}{6G} - \frac{K}{2G a^2} + \frac{4 \pi  \rho_{{\rm m}0 }}{3a^3 } +\frac{4 \pi  \rho_{\gamma 0} }{3a^4 } +\frac{4\pi n_0 N_{\rm f}}{3a^3} \left( \frac{k_0}{a} +  2 G\frac{g(a)}{a} \right) 
\ff
where $\rho_{{\rm m}0}$ and $\rho_{\gamma 0}$ are the present densities of non-relativistic matter and of photons.
We note that the CMC gauge condition is  satisfied in this  case since the momenta are constant densities.  This means that $a$ and $\pi$ are invariant under volume-preserving conformal transformations.

Combining everything, and fixing $N_{\rm lapse}=1$,  we find the modified Friedmann equation
\begin{eqnarray}
\left ( \frac{\dot{a}}{a}  \right )^2 &=& 2GV + \frac{G^2 g^2}{a^2} \nonumber
\\
&=& \frac{\Lambda}{3} -\frac{K}{a^2 } +\frac{8 \pi G \rho_{{\rm m} 0 }}{3 a^3} + \frac{8 \pi  G \rho_{\gamma 0} }{3a^4 }+ \frac{G^2 g^2}{a^2} 
+\frac{8\pi G n_0 N_{\rm f} k_0 }{3a^4}   +  \frac{ 16 \pi  G^2 n_0 N_{\rm f} g(a) }{3a^4}
\label{FRW1}
\end{eqnarray}

\section{Cosmological constraints on time-asymmetric torsion}

\subsection{Massless neutrinos}

We normalize the scale factor to be unity at present. To simplify the argument we assume a simple power-law form $g(a) = C a^{-p}$, though this is not essential for what follows.\footnote{We assume $g(a)$ takes positive values. From the equation of motion for the momentum, Eq.~(\ref{pieom}), we see that positive-valued $g(a)$ represents a friction term, whereas if $g(a)$ is negative it will be a driving term and we'd expect it to cause instabilities.} We find the potential is
\f
V=  \frac{\Lambda}{6G} - \frac{K}{2G a^2} + \frac{4 \pi  \rho_{{\rm m}0 }}{3a^3 } +\frac{4 \pi  \rho_{\gamma 0} }{3a^4 } +\frac{4\pi \alpha  k_0}{3a^4} + \frac{8\pi G  \alpha C}{3a^{4+p}} 
\ff
where $\alpha = n_0 N_{\rm f}$ so the Friedmann equation is 
\f
\left ( \frac{\dot{a}}{a}  \right )^2 = \frac{\Lambda}{3} -\frac{K}{a^2 } +\frac{8 \pi G \rho_{{\rm m}0} }{3 a^3} +\frac{8 \pi G \rho_{\gamma 0} }{3 a^4} + \frac{G^2 C^2}{a^{2+ 2p}} 
+\frac{8\pi G \alpha k_0 }{3 a^4}   +  \frac{ 16\pi  G^2 \alpha C }{3a^{4+p}}
\label{FRW3}
\ff

The modification gives a contribution to the energy density,
\f
\label{newterms}
\rho_{\rm new} =  \left (   \frac{3 }{8 \pi }  \right ) \left [  \frac{G C^2}{a^{2+2p}} +\frac{ 4  G \alpha C}{a^{4+p}}
\right ]
\ff
Let us name the first term in Eq.~(\ref{newterms}) the asymmetric term, since it comes directly from the time-asymmetric modification to the Hamiltonian, and the second term the torsion term, since it comes from the torsion contribution to the connection. We will want to compare this to the usual neutrino energy density (presently assuming massless neutrinos):
\f
\rho_\nu = \frac{2 \alpha k_0}{a^4 }
\ff

The magnitude of the asymmetric term is fixed once the other terms are specified. We start by comparing the various terms at present, $a = 1$. A conservative constraint on the asymmetric term is that it can be no more than the critical density, which in Planck units is of order $10^{-120} m_{\rm Pl}^4$.\footnote{More precisely it is $2 \times 10^{-123} m_{\rm Pl}^4$, but the rounded order of magnitude is sufficient for our purposes.}
In order to estimate the magnitude of the torsion effect, $\alpha C$, we need to know the neutrino number density today. 
The neutrino energy density in the standard cosmology, assuming for now that neutrinos are massless, is $\rho_\nu \simeq 10^{-5}\rho_{\rm crit}$. We saw above that $k_0 \simeq \alpha^{1/3}$, up to a constant of order unity when using natural units.  Hence $C < 10^{-60} m_{\rm Pl}^3$ and $\alpha \simeq (10^{-30} m_{\rm Pl})^3$. Together these ensure that $G\alpha C < 10^{-150} m_{\rm Pl}^4 \simeq 10^{-30} \rho_{\rm crit}\simeq10^{-25}\rho_{\nu}$, ensuring the torsion term in Eq.~(\ref{newterms}) is completely negligible at present.

Given that critical density considerations render the new torsion term unobservable today, we can instead consider the scale dependence of $g(a)$ and investigate possible signatures into the past. Given the scalings, the torsion term is the one that grows fastest into the past for $p$ in the range $0<p<2$. To illustrate the types of scenario that can arise within this range we consider three cases.

\subsubsection*{$p=0$: dark curvature and torsion radiation}

This case has the asymmetric term mimicking curvature with scaling $1/a^2$; such fluids are not useful for matching current observations but are permitted at a subdominant level as their influence becomes less important into the past. Here the torsion term scales the same way as the neutrino kinetic energy and also the photon energy density, so our constraint that it is utterly negligible at the present epoch implies it has been negligible throughout cosmological history.

\subsubsection*{$p=1$: early torsion domination}

The torsion term grows the most rapidly compared to the other terms if we choose $p=1$, whereupon the asymmetric term becomes a contribution to the radiation, as discussed in Ref.~\cite{CGS}, and the new neutrino torsion term grows into the past as $a^5$. Given the powerful constraints we just derived on its magnitude today apply, in order to become significant needs to grow into the past from $G\alpha C \simeq 10^{-30} \rho_{\rm crit}$ to unity. At big bang nucleosynthesis, at redshift $z\simeq 10^{10}$, it will be still negligible compared to standard radiation, $G\alpha C \simeq 10^{-20} \rho_{\rm crit}$. If we continue into the past, at around redshift $z\simeq 10^{30}$, typical of post-inflation reheating epochs, the torsion term becomes comparable to the critical density and could play a role in the reheating mechanism. However, since there are no distinct signatures of reheating models, we still do not gain insight on the form of $g(a)$. We thus conclude that under these assumptions the new torsion term leads to no currently observable cosmological effects. However this scenario is the most promising for a future search for signatures, should the physics of post-inflationary reheating be placed on a firm footing.

\subsubsection*{$p=2$: stiff fluids}

In this case both the asymmetric term and the torsion term scale as $1/a^6$, being the characteristic scaling of a stiff fluid with equation of state $p=\rho$. In this case both terms grow rapidly compared to the matter and radiation terms as we track into the past. yet they must have been small at nucleosynthesis to avoid spoiling it, thus guaranteeing that they are suppressed to at least twenty orders of magnitude below the critical density at present. The stronger constraint on $C$ means that the torsion term is even more negligible compared to the usual neutrino terms at the present epoch.

\subsection{Including neutrino mass}

If we include the effect of neutrino mass $m$, the dispersion relation becomes 
\f
\omega^2 = (k+\mu)^2 + m^2
\ff
The neutrino velocity is 
\f
v \equiv \frac{d\omega}{dk} = \frac{k+\mu}{\sqrt{(k+\mu)^2 + m^2}}
\ff
This has the interesting feature that the neutrinos will be kept relativistic, $v \simeq c$, provided only that $\mu^2 \gg m^2$, regardless of the kinetic energy $k$, i.e.\ the neutrinos would continue to travel at the speed of light even once their kinetic energy fell below the rest mass. However, the above constraint from the Friedmann equation is sufficient to show that $\mu$ is in fact orders of magnitudes below $k$ and hence this regime is not accessed in our Universe.

Inclusion of a small neutrino mass, as motivated by neutrino oscillation experiments, does not alter the argument made above in the massless case, which relied only on the neutrino kinetic energy. Hence we can conclude that the torsion term is still completely negligible in presence of (observationally-permitted) neutrino masses.

\subsection{A speculative scenario}

There is one possible, albeit contrived, 
exception to this. We presume that the asymmetric dark energy term exactly cancels a (negative) bare cosmological constant, i.e.\ in the case $p=-1$ we take $G^2 C^2 =-  \Lambda/3$. Presumably some new fundamental symmetry of nature will have been invoked to justify an exact cancellation. 

In such a scenario both the cosmological constant and the asymmetric dark energy can be orders of magnitude larger than the other terms in the Friedmann and acceleration equations, but if so the constant $C$ is no longer constrained by the critical density as we found above. Taking $C \simeq k_0/G \simeq 10^{-25} m_{\rm Pl}^3$ (i.e.\ over thirty orders of magnitude larger than allowed in the absence of the precise cancellation with $\Lambda$) would allow us to consider the torsion term to be of order of the critical density at the present epoch. In this case, and given that for $p=1$ the torsion term scales likes matter, $1/a^3$, we can propose the torsion term as a dark matter candidate. These background evolution considerations provide both the correct redshift scaling and the correct density today for a dark matter candidate. For this proposal to hold we would next have to derive the first-order perturbation equations and assess the possibility that the new torsion term can clump and match the detailed predictions for the power spectrum of large-scale structure and the cosmic microwave background. 

To examine the implications of this, we consider the fermion contributions to the energy density Eq.~(\ref{hampsi}) brought about by the torsion term.  This is 
\f
\Delta {\cal H}^{\Psi} = -\imath \Pi^A e^a_i \sigma^{ i \ \ B}_A \Delta A_{a B}^C  \Psi_C
\label{hampsi2}
\ff
where
\f
\Delta A_{a B}^C = \frac{2  \imath G}{3} \sigma_{i  B}^C  g(a)
\ff
This gives us
\f
\label{eq35}
\Delta {\cal H}^{\Psi} =  \frac{2 Gg}{a} \Pi^A \Psi_A =\mu (a) \Pi^A \Psi_A 
\ff
This appears to give a new contribution to the energy, which is dependent on the global time, in a way that can be read as 
giving a time-dependent scaling of Newton's constant as experienced by the chiral fermions
\f
G(a) =  G \frac{g}{a}
\ff
As non-chiral particles continue to experience the usual Newton's constant,  this signals a violation of the equivalence principle. Interestingly,  {\it the new term given by Eq.~(\ref{eq35}) is proportional to the 
number density of the chiral fermions.}  This is important because while the baryons dominate the energy density, the neutrinos completely dominate the number density and so would contribute the predominant torsion effect.

\section{Conclusions}

In this article we continued our search for possible observational effects of a class of time-asymmetric modifications of general relativity.  Here we studied the effect of torsion in the equations of motion of neutrinos on the expansion of the universe.  We concluded that in typical cases, and taking into account the number density of cosmic neutrinos and critical-density bounds on the asymmetric term contribution to the Friedmann equation, the additional torsion terms are constrained to be far too small to give rise to observable effects today, and at any recent cosmic epoch. However we found one case with $p=1$ where the torsion term could potentially be significant during post-inflationary reheating.

More speculatively, we uncovered an interesting case in which $p=-1$ and where the asymmetric dark energy is assumed to be cancelled by the bare cosmological constant (including contributions from the vacuum energy of quantum fields).  This torsion term can then be quite large, leading to possible observable effects.  In that case the torsion term scales as a non-relativistic matter contribution.  It turns out to be roughly an enhancement of the energy density proportional to the neutrino number density, mediated by a time-dependent constant.  Given that the neutrino number density dominates over the baryon number density this yields an intriguing, if speculative, possibility for dark matter.

\section*{Acknowledgements}

We would like to thank 
Henrique Gomes
and Vasudev Shyam for discussions. This research was supported in part by Perimeter Institute for Theoretical Physics. Research at Perimeter Institute is supported by the Government of Canada through Industry Canada and by the Province of Ontario through the Ministry of Research and Innovation. This research was also partly supported by grants from NSERC, FQXi and the John Templeton Foundation.  M.C.\ was supported by the EU FP7 grant PIIF-GA-2011-300606 and FCT grant SFRH/BPD/79284/201(Portugal), and  A.R.L.\  by STFC under grant number ST/L000644/1.


\end{document}